\documentclass[12pt]{article}

\begin{document}

\input amssym.tex

\newcommand{\Lor}{L^{\uparrow}_{+}}
\newtheorem{lem}{Lemma}
\newtheorem{defin}{Definition}
\newtheorem{theor}{Theorem}
\newtheorem{prop}{Proposition}
\newtheorem{cor}{Corollary}
\newenvironment{demo}
{\bgroup\par\smallskip\noindent{\it Proof: }}{\rule{0.5em}{0.5em}
\egroup}

\def\Eh{\mbox{\teneufm\char 83}}

\title{Symmetries of the Dirac operators associated with covariantly 
constant Killing-Yano tensors}

\author{Ion I. Cot\u aescu \thanks{E-mail:~~~cota@physics.uvt.ro}\\ 
{\small \it West University of Timi\c soara,}\\
       {\small \it V. P\^ arvan Ave. 4, RO-1900 Timi\c soara, Romania}
\and
Mihai Visinescu \thanks{E-mail:~~~mvisin@theor1.theory.nipne.ro}\\
{\small \it Department of Theoretical Physics,}\\
{\small \it National Institute for Physics and Nuclear Engineering,}\\
{\small \it P.O.Box M.G.-6, Magurele, Bucharest, Romania}}
%\date{\today}
\date{}

\maketitle

\begin{abstract}
The continuous and discrete symmetries of the Dirac-type operators produced 
by particular Killing-Yano tensors are studied in manifolds of arbitrary 
dimensions. The Killing-Yano tensors considered are covariantly 
constant and realize  certain square roots of the metric tensor. Such a 
Killing-Yano tensor produces simultaneously a Dirac-type operator and the 
generator of a one-parameter Lie group connecting this operator with the 
standard Dirac one. The Dirac operators are related among themselves through 
continuous or discrete transformations. It is shown that the groups of the
continuous symmetry  can be only $U(1)$ and $SU(2)$, specific to
(hyper-)K\"ahler spaces, but arising even in cases
when the requirements for these special geometries are not fulfilled. 
The discrete symmetries are also studied obtaining the discrete groups  
${\Bbb Z}_4$ and ${\Bbb Q}$. The briefly presented examples are the 
Euclidean Taub-NUT space and the Minkowski spacetime.

Pacs 04.62.+v

\end{abstract}

\section{Introduction}

The (skew-symmetric) Killing-Yano (K-Y) tensors, that were first introduced 
by Yano \cite{Y} from purely mathematical reasons, are profoundly connected 
to the supersymmetric classical and quantum mechanics on curved manifolds 
where such tensors do exist \cite{GRH}.  The K-Y tensors play an important 
role in theories with spin and especially in the Dirac theory on curved 
spacetimes where they produce first order differential operators, called 
Dirac-type operators, which anticommute with the standard Dirac one, $D$  
\cite{CML}. Another virtue of the K-Y tensors is that they enter as square 
roots in the structure of several second rank St\" ackel-Killing tensors that 
generate conserved quantities in classical mechanics or conserved operators 
which commute with $D$. The construction of Carter and McLenaghan 
depended upon the remarkable fact that the (symmetric) St\" ackel-Killing 
tensor $K_{\mu\nu}$ involved in the constant of motion quadratic in the 
four-momentum $p_\mu$ 
\begin{equation}\label{i1}
Z = {1\over 2} K^{\mu\nu} p_\mu p_\nu
\end{equation}
has a certain square root in terms of K-Y tensors $f_{\mu\nu}$ 
\cite{PF}:
\begin{equation}\label{i2}
K_{\mu\nu} = f_{\mu\lambda} f^{\lambda\,\cdot} _{\cdot\,\nu}\,. 
\end{equation}
\noindent The K-Y tensor here is a $2$-form $f_{\mu\nu} = - f_{\nu\mu}$ 
which satisfies the equation
\begin{equation}\label{i3}
f_{\mu\nu;\lambda}+f_{\mu\lambda;\nu} = 0\,.
\end{equation}

These  attributes of the K-Y tensors lead to an 
efficient mechanism of supersymmetry especially when the 
St\" ackel-Killing tensor $K_{\mu\nu}$ in Eq. (\ref{i1}) is proportional 
with the metric tensor $g_{\mu\nu}$ and the corresponding K-Y tensors 
in Eq. (\ref{i2}) are covariantly constant. Then each tensor of this type, 
$f^i$, gives rise to a Dirac-type operator,  $D^i$,  representing a 
supercharge of the superalgebra $\{ D^i , D^j \} \propto D^2 \delta_{ij} $.

The typical example is the Euclidean Taub-NUT space which is a 
hyper-K\" ahler manifold 
possessing three covariantly constant K-Y tensors with real-valued components 
which constitute a hypercomplex structure generating a $N=4$ superalgebra at 
the level of the Dirac theory \cite{CV0}. Moreover, each involved K-Y tensor 
is a root of the metric tensor as it results from the definition of the 
K\" ahlerian geometries (given in Appendix A). It is worth 
pointing out that the Euclidean Taub-NUT space has, in addition, a 
non-covariantly 
constant K-Y tensor related to its specific hidden symmetry showed off 
by the existence of a conserved Runge-Lenz operator that can be constructed 
with the help of the Dirac-type operators produced by all the four K-Y tensors 
of this space \cite{CV}.   

In what concerns the superalgebras of Dirac-type operators, the inverse 
problem is to find the suitable conjectures allowing the construction of 
Dirac-type operators $D'$ which should satisfy the condition 
$(D')^2\propto D^2$. It 
was shown that these can be produced by covariantly constant K-Y tensors 
having not only real-valued components but also complex ones \cite{K1,K2}. 
This extension seems to be productive since it permits to construct 
superalgebras in the Dirac theory in Minkowski spacetime which is not 
K\" ahlerian, having only complex-valued covariantly constant K-Y tensors 
\cite{K2}. For this reason, in what follows we shall consider such more 
general tensors, called  {\em roots} (instead of complex structures) since all 
of them are, up to constants, roots of the metric tensor. We note that
the complex structures defining K\" ahlerian geometries are 
particular automorphisms of the tangent bundle while the roots we use  
here are automorphisms of the {\em complexified} tangent bundle.   

It is known that in four-dimensional manifolds the standard Dirac operator 
and the Dirac-type ones can be related among themselves through continuous 
or discrete transformations \cite{CV1,K2}. It is interesting that there 
are only two possibilities, namely either transformations of the $U(1)$ group 
associated with the discrete group ${\Bbb Z}_4$ or $SU(2)$ transformations 
and discrete ones of the quaternionic group ${\Bbb Q}$ \cite{CV1,K2}. 
Particularly, in the case when the roots are real-valued (complex structures) 
the first type of symmetry is proper to  K\" ahler manifolds while the second 
largest one is characteristic for hyper-K\" ahler geometries \cite{CV1}. The 
problem is what happens in the case of manifolds with larger number of 
dimensions allowing complex-valued roots. Could we expect to find new larger 
symmetries ? The present article is devoted to this problem.      

Our main purpose is to investigate the specific symmetries of 
the Dirac operators constructed using roots in geometries of arbitrary 
dimensions. We start with the observation that any root give rise 
simultaneously to a Dirac-type operator and a generator of the one-parameter 
Lie group relating this operator to the standard Dirac one. In fact the 
group generator is the main piece of the theory since it is able to produce 
itself the Dirac-type operator through a simple commutation with 
the standard Dirac operator. Exploiting this mechanism we study the 
continuous and discrete symmetries of the Dirac-type operators showing that, 
as in the case of the K\" ahlerian geometries, there exists only two 
types of continuous symmetries, $U(1)$ and $SU(2)$, and the corresponding 
discrete symmetries, ${\Bbb Z}_4$ and ${\Bbb Q}$ respectively. One 
of our important results is the concrete form of the $SU(2)$ transformations 
in the general case of any $4n$-dimensional manifold equipped with 
sets of roots having similar algebraic properties as the quaternion units.       

The paper is organized as follows. We start in the second section with the 
construction of a simple version of the Dirac theory in any dimensions 
defining the Dirac adjoint and the behavior of the Dirac matrices under 
transformations of the universal covering group of the gauge group or of 
the corresponding complexified groups. The next section is devoted to the main 
differential operators we use while in Sec.4 we define the roots and the 
Dirac-type operators discussing their main properties. The continuous 
symmetries of the Dirac-type operators are studied in Sec. 5 and 6 where we 
establish the form of the $U(1)$ and respectively $SU(2)$ transformations 
among the standard Dirac operator and the Dirac-type ones. Therein we 
show that other kind of symmetries are not allowed because of the restrictions 
imposed by the Frobenius theorem. The mentioned discrete symmetries are studied 
in Sec. 7 and finally we present our conclusions and comments. In Appendices 
we give the usual definition of the K\"ahlerian geometries and discuss the 
four-dimensional examples of physical interest, the Euclidean Taub-NUT space 
and the Minkowski spacetime.

\section{Dirac spinors in any dimensions}

The theory of the Dirac spinors in arbitrary dimensions depends on the choice 
of the manifold and the Clifford algebra. Bearing in mind that the irreducible 
representations of the Clifford algebra can have only an odd number of 
dimensions, we start with a $2l+1$-dimensional pseudo-Riemannian manifold 
$M_{2l+1}$ whose flat metric $\tilde\eta$ (of its pseudo-Euclidean model) has 
the signature $( m_{+}, m_{-})$ where $ m_{+} + m_{-}=m=2l+1$. This is the 
maximal manifold that can be associated  to the Clifford algebra  
\cite{Clif} acting on the $2^l$-dimensional space 
$\Psi^{l}$ of the complex spinors $\psi=\varphi_1 \otimes \varphi_2 ...
\otimes\varphi_l$ built using  complex two-dimensional Pauli 
spinors $\varphi$. In this algebra we take the standard Euclidean basis 
formed by the hermitian 
matrices $\tilde\gamma^A= (\tilde\gamma^{A})^{+}$ ($A,B,...=1,2,...,m$) that 
obey  $\{\tilde\gamma^{A},\, \tilde\gamma^{B} \} =2\delta^{A B}I$, and define 
the basis corresponding to the metric $\tilde\eta$ as
\begin{equation}
\gamma^A=\left\{
\begin{array}{lll}
\tilde\gamma^A&{\rm for}& A=1,2,...,  m_{+}\\
i\tilde\gamma^A&{\rm for}& A= m_{+}+1,  m_{+}+2,..., m
\end{array}\right.\,,
\end{equation}
such that
\begin{equation}\label{ACOM}
\{ \gamma^{A},\, \gamma^{B} \} 
=2\tilde\eta^{A B}I\,. 
\end{equation}
Since the first $ m_{+}$ matrices $\gamma^A$ remain hermitian while the 
$ m_{-}$ last ones become anti-hermitian, it seems that the unitaryness 
of the theory is broken. However, this can be restored with the help of the 
generalized  Dirac adjoint,  $\overline{\psi}=\psi^{+}\gamma$, involving a 
hermitian matrix  $\gamma=\gamma^{+}$  which should play the role of a 
metric operator obeying the condition $(\gamma)^2=I$ and giving the Dirac 
adjoint of any matrix $X$ as $\overline{X}=\gamma X^{+} \gamma$. It is not 
difficult to show that the matrix 
$\gamma= \epsilon \,\gamma^{1}\gamma^{2}...\gamma^{m_{+}}$
is convenient for this role \cite{Pro} if we choose the following  values for 
the phase factor:  
\begin{equation}
\epsilon =\left\{
\begin{array}{lll}
(i)^{\frac{{m}_{+}-1}{2}}&{\rm for~ odd}&{m}_{+}<m\\
(i)^{\frac{{m}_{+}}{2}}&{\rm for~ even}&{m}_{+}<m\\
\end{array}\right.\,.
\end{equation}
In the special case of the Euclidean metric (when $ m_{-}=0$) we have 
the trivial solution $\gamma=I$. The algebraic properties of the matrix 
$\gamma$   
depends on  $m_{+}$ such that for $ m_{+}$ taking odd values 
we have  the following superalgebra
\begin{eqnarray}
[\gamma,\,\gamma^{A}]=0& {\rm for}& A=1,2,...,m_{+}\,,
\nonumber\\
\{\gamma,\,\gamma^{A}\}=0& {\rm for}& A=m_{+}+1,...,m\,,
\end{eqnarray}
while for even $ m_{+}$ the situation is reversed. Consequently, one 
can verify that 
\begin{equation}
\overline{\gamma^{A}}=\left\{
\begin{array}{lll}
\gamma^{A}& {\rm for~ odd}& m_{+}\\ 
-\gamma^{A}& {\rm for~ even}& m_{+}   
\end{array}\right.\,,\quad  A=1,2,...,m\,, 
\end{equation}  
which means that from the point of view of the Dirac adjoint all the 
matrices $\gamma^A$ have the same behavior, being either self-adjoint 
or anti self-adjoint. Thus the unitaryness of the theory is guaranteed.
In what follows we consider that $m_{+}$ is odd and match all the phase 
factors according to self-adjoint gamma matrices. When these are anti 
self-adjoint (because of an even $m_{+}$) it suffices to change 
$\gamma^A\to i\gamma^A$.  
Particularly, in the examples we give in different four-dimensional 
manifolds immersed in that of the Kaluza-Klein theory, we take all the  
space-like dimensions of negative signature.

The gauge group $G(\tilde\eta)=O( m_{+}, m_{-})$ of the metric $\tilde\eta$, 
with the mentioned signature, admits an universal covering group 
${\cal G}(\tilde \eta)$ that is simply connected and has the same Lie algebra.
In order to avoid complications due to the presence of these two groups we
consider here that the gauge group is a {\em vector} representation of
${\cal G}(\tilde \eta)$ and we denote by
$G(\tilde \eta)=\lambda[{\cal G}(\tilde \eta)]$
all the equivalent vector representations. The {\em spinor}
representation of the same group, denoted by $\sigma[{\cal G}(\tilde\eta)]$,  
is carried by the space $\Psi^l$ being generated by the spin operators
\begin{equation}\label{SAB} 
S^{AB}=\frac{i}{4}\left[\gamma^{A},\,\gamma^{B}
\right]
\end{equation}
which are self-adjoint, $\overline{S^{AB}} =S^{AB}$,  and satisfy
\begin{eqnarray}
[S^{AB},\,\gamma^{C}]&=&
i(\tilde\eta^{BC}\gamma^{A}-
\tilde\eta^{AC}\gamma^{B})\,,\label{Sgg}\\
{[} S_{AB},\,S_{CD} {]}&=&i(
\tilde\eta_{AD}\,S_{BC}-
\tilde\eta_{AC}\,S_{BD}+
\tilde\eta_{BC}\,S_{AD}-
\tilde\eta_{BD}\,S_{AC})\,.\label{SSS}
\end{eqnarray}
as it results from Eqs. (\ref{ACOM}) and (\ref{SAB}). It is known that for
any real or complex valued skew-symmetric tensor $\omega_{AB}=-\omega_{BA}$ 
the operator 
\begin{equation}\label{TeS}
T(\omega)=e^{iS(\omega)}\,,\quad S(\omega)=\frac{1}{2}
\omega_{AB} S^{AB}
\end{equation}
transforms the gamma-matrices according to the rule
\begin{equation}\label{TgT}
T(\omega)\gamma^{A}[T(\omega)]^{-1}=\Lambda^{A\,\cdot}
_{\cdot\,B}(\omega)\gamma^{B} 
\end{equation}
where
\begin{equation}\label{Lam}
\Lambda^{A\,\cdot}_{\cdot\,B}(\omega)=
\delta^{A}_{B}
+\omega^{A\,\cdot}_{\cdot\,B}
+\frac{1}{2}\,\omega^{A\,\cdot}_{\cdot\,C}
\omega^{C\,\cdot}_{\cdot\,B}+...
+\frac{1}{n!}\,\underbrace{\omega^{A\,\cdot}_{\cdot\,C}
\omega^{C\,\cdot}_{\cdot\, C'}
...\omega^{D\,\cdot}_{\cdot\,B}}_{n}+...\,.
\end{equation}
If $\omega\in {\Bbb R}$ then $\overline{S(\omega)}=S(\omega)$,  the 
operators $T(\omega)$ are unitary with respect to the Dirac adjoint 
satisfying $\overline{T(\omega)}=[T(\omega)]^{-1}$ and, therefore, $T(\omega)
\in \sigma[{\cal G}(\tilde\eta)]$
and $\Lambda(\omega)\in \lambda[{\cal G}(\tilde\eta)]$. However, 
for $\omega \in {\Bbb C}$ we have to consider the {\em complexified}  
group of ${\cal G}$, denoted by ${\cal G}_c$, and the corresponding vector 
and spinor representations.     

\section{Differential operators}

In many physical problems involving  differential operator these are defined 
on a domain of a manifold  which, in general, may be a submanifold of 
$M_{m}$, $m=2l+1$, like in the case of the (1+3)-dimensional spacetimes 
immersed in the five-dimensional manifold of the Kaluza-Klein theory where
$l=2$. Therefore, it is indicated to consider the differential operators on
a submanifold $M_{n}\subset  M_{m}$  of dimension $n\le m$ whose flat metric
$\eta$ is a part (or restriction) of the metric $\tilde\eta$, having the
signature $(n_{+},n_{-})$, with $n_{+}\le  m_{+}$,  $n_{-}\le  m_{-}$ and  
$n_{+}+n_{-}=n$. In $M_{n}$ we choose  a local chart (i.e. natural frame) 
with coordinates $x^{\mu}$, $\alpha,...,\mu,\nu,...=1,2,...,n$, and introduce 
local orthogonal non-holonomic frames using "$n$-beins", $e(x)$ and 
$\hat e(x)$, whose components are labeled by local (hated) indices, 
$\hat\alpha,...\hat\mu,\hat\nu,...=1,2,...,n$, that represent a subset of the  
Latin capital ones, eventually renumbered. The local indices have to be raised 
or lowered by the metric $\eta$. The fields $e$ and $\hat e$ accomplish the 
conditions   
$e_{\hat\alpha}^{\mu}\hat e_{\nu}^{\hat\alpha}=\delta_{\mu}^{\nu}$,
$e_{\hat\alpha}^{\mu}\hat e_{\mu}^{\hat\beta}=\delta_{\hat\alpha}^{\hat\beta}$ 
and  orthogonality relations as 
$g_{\mu\nu} e_{\hat\alpha}^{\mu} e^{\nu}_{\hat\beta}=
\eta_{\hat\alpha \hat\beta}$. With their help   
the metric tensor of $M_n$ can be put in the form   
$g_{\mu\nu}(x)=\eta_{\hat\alpha \hat\beta}
\hat e ^{\hat\alpha}_{\mu}(x) \hat e^{\hat\beta}_{\nu}(x)$.    
The spin connection 
$\Gamma_{\mu}=\frac{i}{2}
e^{\beta}_{\hat\nu}
(\hat e^{\hat\sigma}_{\alpha}\Gamma^{\alpha}_{\beta\mu}-
\hat e^{\hat\sigma}_{\beta,\mu} )
S^{\hat\nu\,\cdot}_{\cdot\,\hat\sigma}$
gives the action of the covariant derivatives in the spinor sector,  
$\nabla_{\mu}\psi=(\tilde\nabla_{\mu}+\Gamma_{\mu})\psi$, where 
$\tilde\nabla_{\mu}$ is the usual covariant derivative (acting in 
natural indices). This definition is 
in accordance with the presented generalized theory of spinors since we have 
$\overline{\Gamma}_{\mu}=-\Gamma_{\mu}$ such that 
the quantity $\overline{\psi}\psi$ can be derived as 
a scalar, i.e. 
$\nabla_{\mu}(\overline{\psi}\psi)=
\overline{\nabla_{\mu}\psi}\,\psi+\overline{\psi}\,\nabla_{\mu}\psi=
\partial_{\mu}(\overline{\psi}\psi)$,
while the quantities $\overline{\psi}\gamma^{\alpha}\gamma^{\beta}...\psi$ 
behave as tensors of different ranks. Moreover, the use of covariant 
derivatives assures the covariance of the whole theory under the  gauge 
transformations, 
$\hat e^{\hat\alpha}_{\mu}(x)\to \hat e'^{\hat\alpha}_{\mu}(x)=
\Lambda^{\hat\alpha\,\cdot}_{\cdot\,\hat\beta}[\omega(x)]
\,\hat e^{\hat\beta}_{\mu}(x)$,
$e_{\hat\alpha}^{\mu}(x)\to  {e'}_{\hat\alpha}^{\mu}(x)=
\Lambda_{\hat\alpha\,\cdot}^{\cdot\,\hat\beta}[\omega(x)]
\,e_{\hat\beta}^{\mu}(x)\label{gauge}$ and
$\psi(x)\to~\psi'(x)=\overline{T}[\omega(x)]\,\psi(x)$
due to $\Lambda[\omega(x)]\in \lambda[{\cal G}(\eta)]$ and 
$T[\omega(x)]\in \sigma[{\cal G}
(\eta)]$ where $\omega_{\hat\mu\hat\nu}=-\omega_{\hat\nu\hat\mu} $ are 
{\em real} functions on $M$. 
We specify that these transformations can not be extended for  $\omega(x)$ 
taking complex values since  $e$ and $\hat e$ must  remain real fields. 

Thus we  reproduced  the main features of the familiar tetrad gauge 
covariant theories with spin in (1+3)-dimensions from which we can take over 
now all the results arising from similar formulas. In this way we find that 
the point-dependent matrices 
$\gamma^{\mu}(x)=e^{\mu}_{\hat\alpha}(x)\gamma^{\hat\alpha}$ and
$S^{\mu\nu}(x)=e^{\mu}_{\hat\alpha}(x)e^{\nu}_{\hat\beta}(x)
S^{\hat\alpha\hat\beta}$
have the same properties as (\ref{ACOM}), (\ref{SAB}), (\ref{Sgg}) and 
(\ref{SSS}), but with $g(x)$ instead of the flat metric, and  recover the 
useful formulas 
\begin{eqnarray}
&&\nabla_{\mu}(\gamma^{\nu}\psi)=\gamma^{\nu}\nabla_{\mu}\psi\,, 
\label{Nabg}\\
&&[\nabla_{\mu},\,\nabla_{\nu}]\psi=
\textstyle{\frac{1}{4}}R_{\alpha\beta \mu\nu}
\gamma^{\alpha}\gamma^{\beta}\psi
\end{eqnarray}
where $R$ is the Riemannian-Christoffel curvature tensor of $M_n$. 

The operators we study here are involving only first order covariant 
derivatives and the natural objects of the theory like the Killing vectors or 
K-Y tensors. The simplest one is the  standard Dirac operator   
\begin{equation}\label{Dirac}
D=i\gamma^{\alpha}\nabla_{\alpha}\,,   
\end{equation}
that is self-adjoint, $\overline{D}=D$, and covariantly transforms 
under gauge transformations, i.e. $D\to D'=\overline{T}DT$. Other operators 
can be constructed using  Killing vectors $\tilde k_{\mu}$ (that obey 
$\tilde k_{\mu;\nu}+\tilde k_{\nu;\mu}=0$) generalizing the result of Carter 
and McLenaghan \cite{CML} obtained for the Dirac fermions. Thus one can show 
that for any 
isometry of $M_n$ of Killing vector $\tilde k^\mu$ there is an appropriate 
operator 
\begin{equation}\label{kil}
X_{\tilde k} = i \left(\tilde k^\mu \nabla_\mu - \textstyle{1\over 4} 
\gamma^\mu \gamma^\nu \tilde k_{\mu;\nu}\right)
\end{equation}
which {\em commutes} with the standard Dirac operator. In addition, following 
the arguments of Ref. \cite{ES} one finds that $X_v$ is just a generator of 
the representation of the universal covering group of the {\em isometry} 
group of $M_n$ induced by $\sigma[{\cal G}(\eta)]$. Another result of 
Refs. \cite{CML, MLS, KML} which holds in the $n$-dimensional case is that
each K-Y tensor $\tilde f$ of rank 2 produces the non-standard Dirac operator 
of the form 
\begin{equation}\label{df}
D_{\tilde f} = i\gamma^\mu \left(\tilde f_{\mu\,\cdot}^{\cdot\,\nu}\nabla_\nu  
- \textstyle{1\over 6}\gamma^\nu 
\gamma^\rho \tilde f_{\mu\nu;\rho}\right)
\end{equation}
which {\em anticommutes} with  $D$. The operators $D_{\tilde f}$ 
will be called Dirac-type operators. 
In the following we shall focus on the properties of these  operators 
and other ones related to  K-Y tensors. 

\section{Roots and Dirac-type operators}

Given $\xi$ an arbitrary tensor field of rank 2  defined on a domain of $M_n$, 
we denote with the {\em same} symbol $\left<\xi\right>$ the equivalent
matrices with the elements $\xi^{\mu\,\cdot}_{\cdot\,\nu}$ in natural
frames as well as those having the matrix elements
$\xi^{\hat\alpha\,\cdot}_{\cdot\,\hat\beta}=\hat e_{\mu}^{\hat\alpha}
\xi^{\mu\,\cdot}_{\cdot\, \nu}e^{\nu}_{\hat\beta}$
in local frames. We say that $\xi$ is non-singular on
$M_n$ if det$\,\left<\xi\right>\not=0$  on a domain of $M_n$ where the metric 
is non-singular. This tensor is said irreducible on $M_n$ if its matrix is 
irreducible. 
\begin{defin}\label{Def1} 
The  non-singular real or complex-valued K-Y tensor $f$ of rank 2 defined on 
$M_n$ which satisfies
\begin{equation}\label{fi}
f^{\mu\,\cdot}_{\cdot\,\alpha} f_{\mu\beta}=g_{\alpha\beta}\,,
\end{equation}
is called an unit root of the metric tensor of $M_n$, or simply an unit root 
of $M_n$.  
\end{defin}
Any K-Y tensor that satisfy Eq. (\ref{fi}) is covariantly constant \cite{K1}, 
i.e.,
\begin{equation}\label{cc}
f_{\mu\nu;\sigma}=0\,.
\end{equation}  
Since Eq. (\ref{fi}) can be written as
$f^{\mu\,\cdot}_{\cdot\,\alpha}f^{\alpha\,\cdot}_{\cdot\,\nu}=
-\delta^{\mu}_{\nu}$ 
this takes the matrix form,   
\begin{equation}\label{f2I}
\left<f\right>^2=-{\bf  I}\,,
\end{equation}
where the notation ${\bf I}$ stands for the $n\times n$ identity matrix.
Hereby we see that the unit roots are matrix representations of several 
{\em complex units} (similar to $i\in {\Bbb C}$) with usual properties as,
for example, $\left<f\right>^{-1}=-\left<f\right>$. The unit roots having
only {\em real}-valued components are called {\em complex structures}  
and represent automorphisms of the tangent fiber bundle ${\cal T}(M_n)$ of 
$M_n$. In local frames these appear as particular point-dependent
transformations of the gauge group $G(\eta)=\lambda[{\cal G}(\eta)]$. 
The manifold possessing such structures are said to have a K\" ahlerian 
geometry (see the Appendix A). However, the unit roots considered here 
are beyond this case since these are defined as automorphisms of the 
complexified fiber bundle ${\cal T}(M_n)\otimes {\Bbb C}$, being thus 
transformations of the complexified group 
$G_c(\eta)=\lambda[{\cal G}_c(\eta)]$.   

As in the case of the complex structures of the K\" ahlerian geometries, the 
matrices of the unit roots have specific algebraic properties resulted from
Eq. (\ref{f2I}). These can be pointed out in local frames (where the matrix
elements are $f^{\hat\alpha\,\cdot}_{\cdot\,\hat\beta}=
\hat e_{\mu}^{\hat\alpha} f^{\mu\,\cdot}_{\cdot\, \nu} e^{\nu}_{\hat\beta}$)
using gauge transformations of $G(\eta)$.
\begin{lem}
The matrix  of any root of $M_n$ is equivalent with a matrix completely
reducible in $2\times 2$ diagonal blocks.
\end{lem}
\begin{demo}
The matrix $\left<f\right>$ which satisfies Eq. (\ref{f2I}) has only
two-dimensional invariant
subspaces spanned by pairs of vectors $z$ and ${\left<f\right>}z$.
On these subspaces, Eq. (\ref{f2I}) is solved in local frames by two types
of $2\times 2$ unimodular blocks without diagonal elements: either
skew-symmetric blocks with factors $\pm 1$, when the involved dimensions are
of the same signature, or symmetric ones with pure imaginary phase factors,
$\pm i$, if the signatures are opposite. However, the diagonalization
procedure can not be continued using transformations of $G(\eta)$ since
these preserve the form of the $2\times 2$ blocks which are proportional with
the generators of the subgroups $SO(2)$ or $SO(1,1)$ acting on the
corresponding invariant subspaces. Notice that other transformations of
$G_c(\eta)$ are not useful since these do not leave the metric invariant.
\end{demo}\\
This selects the geometries allowing unit roots.
\begin{cor}
The unit roots are allowed only by manifolds $M_n$ with an even number of
dimensions, $n=2k, \, k\le l$.
\end{cor}
\begin{demo}
If $n$ is odd then the $2\times 2$ blocks do not cover all dimensions, so that
${\rm det}\hat{\left<f\right>}=0$ and $f$ is no more an unit root of $M_n$.
\end{demo}
\begin{cor}\label{coco}
The unit roots of $M_n$ have real matrices only when the metric $\eta$ 
has a signature with even $n_{+}$ and $n_{-}$.  
Otherwise the unit roots have only complex-valued matrices. 
In both cases the matrices of the unit roots are unimodular, 
i.e. {\rm det}$\left<f\right>=1$. 
\end{cor} 
It is clear that for the real-valued unit roots (i.e., complex structures)
one can construct the {\em symplectic} 2-forms
$\tilde\omega =\frac{1}{2}f_{\mu\nu}dx^{\mu}\land dx^{\nu}$ which are
closed and  non-degenerate.

The above properties indicate that the unit roots are defined up to
sign. Therefore, if two unit roots $f_1$ and $f_2$ do not obey the
condition $f_1=\pm f_2$ then these will be considered {\em different}
between themselves. We denote by ${\cal R}_1(M_n)$ the set of all different
unit roots of the manifold $M_n$. On the other hand, when an unit root $f$
is multiplied by an arbitrary {\em real} number
$\alpha \not=0$, we say that $\xi(x)=\alpha f(x)$ is a {\em root} of norm 
$\|\xi\|=|\alpha|$. Thus we can associate to any unit root $f$ 
the one-dimensional linear real space $L_f =\{\xi\,|\,\xi=\alpha f,\, 
\alpha\in {\Bbb R}\}$ in which each  non vanishing element is a root.
According to Corollary \ref{coco}, when the metric $\eta$ is pseudo-Euclidean,
the unit roots can have complex matrix elements and in that case the unit
root $f$ and its {\em adjoint}, $f^*$, are different. This last one generates
its own linear real space $L_{f^*}$  of adjoint roots  which satisfy 
$\left[ \,\left<\xi \right>^* ,\,\left<\xi'\right>\,\right] =0,\, \forall 
\xi,\, \xi' \in L_f$ since the matrices of $f$ and $f^*$ commutes with each 
other, having same diagonal blocks up to signs.

The whole set of roots of $M_n$ defined as
\begin{equation}\label{setR}
{\cal R}(M_n)=\bigcup_{f\in {\cal R}_1(M_n)}(L_f-\{0\})
\end{equation}
seems to have special algebraic structure since it does not have the
element zero and, in general, it is not certain that a linear
combination of roots is a root too. To convince this it is enough to observe 
that the sum of the roots $\xi$ and  $\xi^*$ is no more a root since 
det$(\left<\xi\right>+\left<\xi\right>^*)=0$ when $\xi^*\not=\xi$ because
of the reduction of the pure imaginary diagonal $2\times 2$ blocks. Moreover,
the product of the matrices of two different roots gives a nonsingular matrix 
but that may be not of a root. Thus we understand that ${\cal R}(M_n)+\{0\}$
can not be organized as a global linear space or algebra even though,
according to the definition (\ref{setR}), it naturally includes linear parts
as  $L_f$ or $L_{f^*}$. In other respects, we know examples  
indicating that ${\cal R}(M_n)$  may contain subsets which are parts of some 
linear spaces with one or three dimensions, isomorphic with Lie algebras 
\cite{CV1,K2}. In any event, the algebraic properties and the topology of
${\cal R}(M_n)$ seem to be complicated depending on the
topological structure of the set of unit roots ${\cal R}_1(M_n)\subset 
{\cal R}(M_n)$. 

The K-Y tensor gives rise to  Dirac-type operators of the form 
(\ref{df}). These have  an important property formulated in the next theorem 
\cite{K1}.
\begin{theor}\label{D2D}
The Dirac-type operator $D_f$ produced by the K-Y tensor $f$  
satisfies the  condition  
\begin{equation}\label{D2D2}
(D_{f})^2=D^2 
\end{equation}
if and only if $f$ is an unit root. 
\end{theor}
\begin{demo}
The arguments of Ref. \cite{K1} show that the condition Eq. (\ref{D2D2}) is 
equivalent  with Eqs. (\ref{fi}) and (\ref{cc}). 
Moreover, we note that for $f\in {\cal R}_1(M_n)$  the square of 
the Dirac-type operator
\begin{equation}\label{Dirf}
D_f=if_{\mu\,\cdot}^{\cdot\,\nu}\gamma^{\mu}\nabla_{\nu}
\end{equation}
has to be calculated exploiting  the identity 
$0=f_{\mu\nu;\alpha;\beta} -f_{\mu\nu;\beta;\alpha}=
f_{\mu\sigma} R^{\sigma}_{\cdot\,\nu\alpha\beta}+
f_{\sigma\nu} R^{\sigma}_{\cdot\,\mu\alpha\beta}$,
which gives  
$R_{\mu\nu\alpha\beta}f^{\mu\,\cdot}_{\cdot\,\sigma} f^{\nu\,\cdot}_{\cdot\,
\tau}
=R_{\sigma\tau\alpha\beta}$ and leads to Eq. (\ref{D2D2}). 
\end{demo}\\
Thus we  conclude that the equivalence of the condition (\ref{D2D2}) 
with Eqs. (\ref{fi}) and (\ref{cc}) holds in any geometry of dimension $n=2k$ 
allowing roots. When $f^*\not=f$ then $D_{f^*}$ is different from $D_f$ even 
if $(D_f)^2=(D_{f^*})^2=D^2$. These operators are no longer self-adjoint, 
$\overline{D_f}=D_{f^*}$, obey
\begin{equation}\label{DDO}
\left\{ D_{f},\,D\right\}=0\,,\quad
\left\{ D_{f^*},\,D\right\}=0\,,\quad
\end{equation}
and commute with each other.  

\section{Continuous symmetries generated by unit roots} 

Thus we arrive at the main point of our study, namely to find the continuous 
transformations able to relate the operators $D_f$ and $D$ to each other. 
We know that in many particular cases \cite{CV1, K2, CV} this is possible 
and now we intend to point out that this is a general property of theories 
involving roots. To this end we introduce a new useful point-dependent matrix.    
\begin{defin}
Given the unit root $f$, the matrix 
\begin{equation}\label{SfS}
\Sigma_{f}=\frac{1}{2}f_{\mu\nu}S^{\mu\nu}
\end{equation}
is the spin-like operator associated to $f$.
\end{defin}
This is an operator that acts on the space of spinors $\Psi^k$ and, therefore, 
can be interpreted as a generator of the spinor representation 
$\sigma[{\cal G}_c(\eta)]$ since the components of $f$ are, in general, 
complex-valued functions.  It has the obvious property $\overline{\Sigma_{f}}= 
\Sigma_{f^*}$ while from (\ref{Nabg}) and (\ref{fi}) one 
obtains that it is covariantly constant in the sense that
$\nabla_{\nu}(\Sigma_f\psi)=\Sigma_f \nabla_{\nu} \psi$.   
Hereby we find that the Dirac-type operator (\ref{Dirf}) can be written as
\begin{equation}\label{DDS}
D_f=i\left[D,\, \Sigma_f\right]
\end{equation}
where $D$ is the standard Dirac operator defined by Eq. (\ref{Dirac}).
Moreover, from Eqs. (\ref{DDO}) we deduce 
$[\Sigma_f,\, D^2]=[\Sigma_f,\,(D_{f})^2]=0$ 
and similarly for $\Sigma_{f^*}$.

\begin{defin}
For each $f\in {\cal R}_1(M_n)$ we define its associated one-parameter Lie 
group ${\cal G}_{f} \subset {\cal G}_c$ as the group whose operators in 
the spinor representation are
\begin{equation}\label{Taf}
T(\alpha f)=e^{i\alpha \Sigma_f} \in \sigma({\cal G}_f)
\end{equation}
where  $\alpha \in{\Bbb R}$ is the group parameter.  
\end{defin}  
In other words these are operators of the form $(\ref{TeS})$ 
where we replace $\omega$ by roots $\xi=\alpha f\in L_f$. Consequently, their 
action 
on the point-dependent Dirac matrices results from  Eq. (\ref{TgT}) to be,    
\begin{equation}\label{TgT1}
T(\alpha f)\gamma^{\mu}[T(\alpha f)]^{-1}=
\Lambda^{\mu\,\cdot}_{\cdot\,\nu}(\alpha f)\gamma^{\nu}\,, 
\end{equation}
where  
$\Lambda^{\mu\,\cdot}_{\cdot\,\nu}= 
e^{\mu}_{\hat\alpha}\Lambda^{\hat\alpha\,\cdot}_{\cdot\,\hat\beta}
\hat e^{\hat\beta}_{\nu}$ are matrix elements with natural indices of the 
matrix 
\begin{equation}\label{Lamaf}
\Lambda(\alpha f)=e^{\alpha \left<f\right>}={\bf I}\cos\alpha+ 
\left<f\right>\sin\alpha  
\end{equation}   
calculated according to Eqs. (\ref{Lam})  and (\ref{f2I}). We note that this 
equation is a matrix representation of the usual {\em Euler formula} of the 
complex numbers. Now it is obvious that in local frames
$\left<f\right>=\Lambda(\frac{\pi}{2}f)\in \lambda({\cal G}_c)$, as mentioned
above.
\begin{theor}\label{TDT}
The operators  $T(\alpha f)\in \sigma({\cal G}_f)$ have the following action
in the space of the operators $D$ and $D_f$:
\begin{eqnarray}
T(\alpha f) D [T(\alpha f)]^{-1}&=&D\cos\alpha-D_f\sin\alpha\,,\label{T1}\\        
T(\alpha f) D_f [T(\alpha f)]^{-1}&=&D\sin\alpha+D_f\cos\alpha\,.        
\label{T2}
\end{eqnarray}
\end{theor}
\begin{demo}
From Eq. (\ref{Lamaf}) we obtain the matrix elements
$\Lambda^{\mu\,\cdot}_{\cdot\,\nu}(\alpha f)=
\cos\alpha\, \delta^{\mu}_{\nu}+\sin\alpha\,  
f^{\mu\,\cdot}_{\cdot\,\nu}$
which lead to the desired result because $\Sigma_f$ is covariantly constant. 
\end{demo}\\
From this theorem it results that $\alpha\in [0,2\pi]$ and, consequently, 
the group ${\cal G}_f\sim U(1)$ is {\em compact}. Therefore, it must be  
subgroup of the maximal compact subgroup of ${\cal G}_c$. In addition, 
from Eq. (\ref{Lamaf}) we see that $L_{f}\sim so(2)$ is the Lie algebra of the 
vector representation of ${\cal G}_f$ that is the compact group 
$\lambda({\cal G}_f)=\{\Lambda(\alpha f)|\alpha\in [0,2\pi]\} \sim SO(2)$. 
Note that the transformations (\ref{T1}) and (\ref{T2})  
leave invariant the operator $D^2=(D_f)^2$  because this commutes 
with the spin-like operator $\Sigma_f$ which generates these transformations.  

Particularly, if $M_n$ allows real-valued unit roots (i.e. complex structures) 
this is an usual K\" ahler manifold. In general, when $f$ has complex 
components (and $f^*\not=f$) then $L_{f^*}\sim so(2)$ is a different linear 
space representing the Lie algebra of $\lambda({\cal G}_{f^*})$. These two Lie 
algebras are complex conjugated to each other but remain isomorphic since 
they are real algebras.  
The relation among the transformations of $\sigma({\cal G}_{f})$ and 
$\sigma({\cal G}_{f^*})$ is
$\overline{T}(\alpha f)=T(-\alpha f^*)=[T(\alpha f^*)]^{-1}$ 
which means that when $f^*\not=f$ the representation $\sigma({\cal G}_f)$ is 
no more unitary in the sense of the generalized Dirac adjoint. 

The conclusion is that an unit root produces simultaneously a Dirac-type 
operator $D_f$ which satisfies Eq. (\ref{D2D2}) and the one-parameter  
Lie group ${\cal G}_f$ one needs to relate $D$ and $D_f$ to each 
other.  
   
\section{Symmetries due to families of unit roots}

The next step is to investigate if there could appear higher symmetries 
given by non-abelian Lie groups with many parameters embedding different 
abelian groups ${\cal G}_f$ produced by some sets of unit roots which have to 
form bases of linear spaces isomorphic with the Lie algebras of these 
non-abelian groups. Such Lie algebras must include many one-dimensional Lie 
algebras $L_f$ being thus subsets of ${\cal R}(M_n)+\{0\}$ where we know that 
the linear properties are rather exceptions. Therefore, we must look for 
special {\em families} of unit roots, $F=\{f^i\,|\,i=1,2,...,N_F\}\subset 
{\cal R}_1(M_n)$, having supplementary  properties which should guarantee 
simultaneously that: 
(I) the linear space $L_{F}=\{\xi\,|\,\xi=\xi_i f^i,\, \xi_i\in {\Bbb R}\}$ 
is isomorphic with a real Lie algebra, and (II) each element 
$\xi\in L_{F}-\{0\}$ is a root (of an arbitrary norm). 

The first condition is accomplished only if the set 
$\{ T(\xi)\,|\, \xi \in L_F\}$ 
includes a Lie group with $N_F$ parameters. This means that the  operators 
$\Sigma^i=\Sigma_{f^i},\, i=1,2,...N_F $ must be (up to constant factors) the 
basis-generators of a Lie algebra with some real structure constants
$c_{ijk}$. Then according to Eqs. (\ref{SSS}) and (\ref{SfS}), we can write
\begin{equation}\label{SSffS}
\left[\Sigma^i, \Sigma^j\right]=
\textstyle{\frac{i}{2} \left[\,\left<f^i\right>,\,\left<f^j\right>\,
\right]_{\mu\nu}} S^{\mu\nu}= ic_{ijk}\Sigma^k 
\end{equation}
obtaining a necessary condition for $F$ be a family of unit roots,
\begin{equation}\label{ffcf}
{\textstyle [\,\left<f^i\right>,\,\left<f^j\right>\, ]}=c_{ijk}
{\textstyle \left< \right.} f^k {\textstyle \left. \right>}\,.
\end{equation} 
The condition (II) is accomplished only when $\left<\xi\right>^2$ is equal up 
to a 
positive factor (i.e. the squared norm) with $-{\bf I}$. This requires to 
have
\begin{equation}\label{ffkf}
 \left\{\,\textstyle{\left<f^i\right>,\,\left<f^j\right>}\,\right\}
=-2\kappa_{ij} {\bf I}\,.
\end{equation} 
where $\kappa$ is a {\em positive definite} metric that can be brought in 
canonical form $\kappa_{ij}=\delta_{ij}$ through a suitable choice of the 
unit roots.  
If $F$ satisfy simultaneously Eqs. (\ref{ffcf}) and (\ref{ffkf}) then $L_F$ is 
just the Lie algebra of the group $\{\Lambda(\xi)\,|\, \xi\in L_F\}$ the 
matrices of which read
\begin{equation}\label{LX}  
\Lambda(\xi)=\textstyle{e^{\xi_i \left<f^i\right>}={\bf I}\cos\|\xi\|+
\nu_i \left<f^i\right>}\sin\|\xi\|
\end{equation} 
where $\|\xi\|=\sqrt{\xi_i\xi_i}$ (when we take $\kappa_{ij}=\delta_{ij}$) and 
$\nu_i =\xi_i/\|\xi\|$. All these results lead to the following theorem.
\begin{theor}
If the set $F=\{f^i\,|\,i=1,2,...,N_F\} \in {\cal R}_1(M_n)$ is a family of 
unit roots then the matrices ${\bf I}$ and $\left<f^i\right>,\,
i=1,2,...,N_F,$ form the basis of a matrix representation of a
finite-dimensional associative algebra over ${\Bbb R}$.      
\end{theor}
\begin{demo}
If $F$ is a family of unit roots in the sense of above definition then 
$f^i$ must satisfy Eqs. (\ref{ffcf}) and (\ref{ffkf}) with canonical metric.
Hereby it results that the set of the real linear combinations
$\xi_{0}{\bf I}+\xi_{i}\left<f^i\right>$ forms an associative algebra.
This algebra is closed with respect to the matrix multiplication 
that can be calculated by adding the commutator and anticommutator. 
Moreover this algebra is a division one. There exists the zero element
(with $\xi_0=0,\,\xi_i=0$), the unit element is ${\bf I}$ and each element
different from zero has the inverse  
$(\xi_{0}{\bf I}+\xi_{i}\left<f^i\right>)^{-1}=
(\xi_{0}{\bf I}-\xi_{i}\left<f^i\right>)
/({\xi_0}^2+\xi_i\xi_i)$. 
Obviously, this real algebra is finite possessing a basis of dimension
$N_F+1$ where $\left<f^i\right>$ play the role of complex
units. Eq. (\ref{LX}) can be interpreted as a matrix 
representation of the Euler formula.
\end{demo}\\
This theorem severely restricts the  existence of the families of unit roots. 
Indeed, according to the Frobenius theorem there are only two finite real 
algebras able to give suitable representations in spaces of roots, namely the 
algebra ${\Bbb C}$ of complex numbers and the {\em quaternion} algebra, 
${\Bbb H}$. In the first case we have {\em isolated} unit roots $f$  and 
representations of the ${\Bbb C}$ algebra generated by the matrices ${\bf I}$ 
and $\left<f\right>$ (which play the role of $i\in {\Bbb C}$) related to the 
continuous symmetry  group ${\cal G}_f\sim U(1)$ we studied in the previous 
section. 

Here we focus on the second possibility leading to families of unit roots with 
$N_F=3$ that constitute matrix representations of the quaternion units. 
\begin{theor}\label{bumbum}
The unique type of family of unit roots with $N_F>1$ having the properties (I) 
and (II) are the triplets $F=\{ f^1,f^2,f^3\}\subset {\cal R}_1(M_n)$ which 
satisfy 
\begin{equation}\label{algf}
{{\textstyle  \left<f^i\right>\,\left<f^j\right>}=-\delta_{ij} {\bf  I}} +
\varepsilon_{ijk} {\textstyle  
\left< \right.} f^k {\left. \textstyle \right>}\,, 
\quad i,j,k...=1,2,3\,. 
\end{equation}
\end{theor}
\begin{demo}
Taking into account that $\varepsilon_{ijk}$ is the anti-symmetric tensor 
with $\varepsilon_{123}=1$ we recognize that Eqs. (\ref{algf}) are the 
well-known multiplication rules of the quaternion units or similar algebraic 
structures (e.g. the Pauli matrices). Consequently, the matrices  
$\left<f^i\right>$ and ${\bf I} $ generate a matrix representation of 
${\Bbb H}$. The Frobenius theorem forbids other choices.
\end{demo}\\
If the unit roots $f^i$ have only real-valued components 
we recover the {\em hypercomplex structures} defining hyper-K\" ahler 
geometries (presented in the Appendix A). 

Eqs. (\ref{algf}) combined with the previous results (\ref{SSffS})-(\ref{LX}) 
provide all the features of the specific continuous symmetry associated to 
$F$. First we observe that the symmetry is governed by the group 
${\cal G}_F\sim SU(2)\subset {\cal G}_c(\eta)$ whose vector representation is 
the compact group $\lambda({\cal G}_F)=
\{\Lambda(\xi)\,|\,\xi\in L_F,\, \|\xi\|\le 2\pi\}$ formed by the 
matrices (\ref{LX}) constructed using the elements of the Lie algebra 
$L_F\sim su(2)\sim so(3)$.
\begin{cor}
The basis-generators of $L_F$ are $\frac{i}{2} f^i$ while the basis-generators 
of the $su(2)$ algebra of ${\cal G}_F$ in spinor representation read 
${\cal S}^i=\frac{1}{2}\Sigma^i$ $(i=1,2,3)$. 
\end{cor}
\begin{demo}
From Eqs. (\ref{algf}) and (\ref{ffcf}) we deduce that 
$c_{ijk}=2\varepsilon_{ijk}$. 
Furthermore, from  Eqs. (\ref{SSffS}) and (\ref{ffcf})  
we obtain the standard commutation rules of $SU(2)$ generators, 
\begin{equation}
\left[ {\cal S}^i,\,{\cal S}^j \right]=i\varepsilon_{ijk}{\cal S}^k\,,
\end{equation}  
and similarly for  $\frac{i}{2}\left<f^i\right>$.
\end{demo}\\
We specify that the operators of $\sigma({\cal G}_F)$, 
\begin{equation}\label{TxiS}
T(\xi)=e^{i\xi_i \Sigma^i} =e^{2i\xi_i {\cal S}^i}\,, 
\quad \xi=\xi_i f^i\in L_F, 
\end{equation}
can be calculated directly from Eq. (\ref{Taf}) replacing 
$\alpha=\pm\|\xi\|=\pm\sqrt{\xi_i\xi_i}$ 
and $f=\pm \xi/ \|\xi\|$. Then the transformations the Dirac
matrices (\ref{TgT1}) can be expressed in terms of the parameters $\xi_i$ 
using the matrices $\Lambda(\xi)$ of the form (\ref{LX}). Hereby we 
observe that $\xi_i$ are nothing other than the analogous of the well-known 
Caley-Klein parameters but ranging in a larger spherical domain (where 
$\|\xi\|\le 2\pi$) such that they cover two times the group 
${\cal G}_F\sim SU(2)$.  

In the case of triplets involving only real-valued unit roots when the 
geometry is  hyper-K\" ahler, each family of real unit roots (i.e., a  
hypercomplex structure) $F$ has its own Lie algebra $L_F\sim su(2)$. These 
algebras can not be embedded in a larger one because of the restrictions 
imposed by the Frobenius theorem.  An example of hyper-K\" ahler manifold 
is the Euclidean Taub-NUT space which is equipped with only one family of 
real unit roots \cite{CV, CV1} (see the Appendix B). The manifolds with 
pseudo-Euclidean metric with odd $n_{+}$ and $n_{-}$ have only pairs of 
{\em adjoint} triplets, $F$ and $F^*$, the last one being formed by the 
adjoints of the unit roots of $F$. The spaces 
$L_F$ and $L_{F^*}$ are isomorphic between themselves (as real vector spaces) 
and all the results concerning the symmetries generated by $F^*$ can be taken 
over from those of $F$ using complex conjugation. Moreover, we must specify 
that the set $L_{F}\bigcup L_{F^*}$ is no more a linear space since the linear 
operations among the elements of $L_F$ and $L_{F^*}$ are not allowed. A typical 
example is the Minkowski spacetime which has a pair of conjugated triplets of 
complex-valued unit roots \cite{K2} (presented in Appendix C). Both these 
examples of manifolds possessing triplets with the properties (\ref{algf}) are 
of dimension four. As we know, the results indicate that similar properties 
could 
have all the flat manifolds of dimension $n=4k,\, k=1,2,3,...$ where 
we expect 
to find many such triplets \cite{GM}. 

Starting with a triplet $F=\{f^1,f^2,f^3\}\subset {\cal R}_1(M_n)$ with 
the properties (\ref{algf}) one can construct a rich set of Dirac-type 
operators of the form $D_{\nu}=\nu_i D^i$ where $\nu$ is an unit vector 
(with $\nu_i\nu_i=1$) and $D^i= D_{f^i}=i[D,\,\Sigma^i]$, $i=1,2,3$, play the 
role of a {\em basis}. This set is {\em compact} and isomorphic with the 
sphere of unit roots ${\bf S}_{F}=\{f_{\nu}\,|\,f_{\nu}=\nu_i f^i,\, \nu^2=1\}
\subset L_F$, since $D_{\nu}=D_{f_{\nu}}$ for any $f_{\nu}\in {\bf S}_F$. 
Each operator $D_{\nu}$ can be related to $D$ through the transformations  
(\ref{T1}) and (\ref{T2}) of the spinor representation of the one-parameter 
subgroup ${\cal G}_{f_{\nu}}\subset {\cal G}_F \sim SU(2)$  defined by 
$f_{\nu}$. The general $SU(2)$ transformations in the basis $D$, $D^i$ 
($i=1,2,3$) read
\begin{eqnarray}  
T(\xi)D[T(\xi)]^{-1}&=&D\cos \|\xi\|-\nu_i D^i\sin\|\xi\|\,,\label{TDT10}\\ 
T(\xi)D^i[T(\xi)]^{-1}&=&D^i\cos \|\xi\|+ 
(\nu_i D+\varepsilon_{ijk} \nu_j D^k)\sin\|\xi\|\,, \label{TDT20}
\end{eqnarray}
where $\xi=\|\xi\| \nu_i f^i$. 

The basis operators $D^i$ anticommute with $D$ and, consequently, all the 
Dirac-type operators $D_{\nu}$ anticommute with $D$. Moreover, the next 
theorem extends this property up to a $N=4$ superalgebra. 
\begin{theor}
If a triplet $F\subset {\cal R}_1(M_n)$ accomplishes  Eqs. (\ref{algf}) then 
the corresponding Dirac-type operators satisfy the superalgebra
\begin{equation}\label{4sup}
\left\{D^i,\,D^j\right\}=2\delta_{ij} D^2\,. 
\end{equation}
\end{theor}
\begin{demo}
If $i=j$ we take over the result of Theorem \ref{D2D}. For $i\not=j$ the 
concrete calculation shows that $D^i$ and $D^j$ anticommute. 
 \end{demo}\\
Thus it is clear that the operators $D$ and $D^i$ ($i=1,2,3$) form a 
superalgebra with four supercharges \cite{CV}. Moreover, Eqs. (\ref{4sup}) 
remain invariant under transformations (\ref{TDT10}) and (\ref{TDT20}) 
since $T(\xi)$ and $D^2$ commute between themselves. 

The previous results indicate that the set ${\cal R}_1(M_n)$, of unit roots 
producing Dirac-type operators, has an interesting topological structure 
involving either single $f$  producing 
isolated  Dirac-type operators  or unit spheres  ${\bf S}_{F}$ leading to 
compact sets of Dirac-type operators. In order to show off this structure one 
needs to correctly identify all the independent triplets with the properties 
(\ref{algf}) or the different singles. To this end one has to exploit 
the mechanisms of our theory based on the fact that the linear spaces $L_f$ or 
$L_F$ are isomorphic with the Lie algebras of the symmetry  groups of the  
Dirac-type operators generated by spin-like operators.

\section{Discrete symmetries}

In the case when there appear difficulties because of a large number of 
different singles or triplets, the  study of the discrete symmetries could 
be also productive. Of course, the results concerning the continuous symmetries 
obtained above will be crucial for understanding the structure of the discrete 
transformations which relate among themselves the standard Dirac operator and 
the Dirac-type ones.

Let us start with the simplest case of an isolated unit root $f$. 
\begin{theor}\label{D1}
For any unit root $f$ there exists the discrete group 
${\Bbb  Z}_4(f)\subset {\cal G}_f$ the orbit of which is $\{D,-D,D_f,-D_f\}$.
\end{theor}
\begin{demo}
Using Eqs. (\ref{T1}) and (\ref{T2}) one observes that the transformations 
$I$, $U_f=T(-\frac{\pi}{2}f)$, $P=(U_f)^2=T(\pi f)$, and 
$(U_f)^3 = T(\frac{\pi}{2}f)=PU_f=U_f P$ form the spinor representation of 
the cyclic group ${\Bbb  Z}_4(f)$. Since  $P^2=I$, the pair $(I, P)$ represents 
the subgroup ${\Bbb  Z}_2\subset {\Bbb  Z}_4(f)$. According to Eq. (\ref{T1}) 
we find that    
\begin{equation}\label{DUDU}
D_f=U_f D(U_f)^{-1}
\end{equation}
and the action of the operator $P$, 
\begin{equation}\label{pari}
P\gamma^{\mu}P=-\gamma^{\mu}\,,  
\end{equation}
is independent on the form of $f$ so that this changes the sign of all the 
Dirac or Dirac-type  operators.  
\end{demo}\\ 
For a given manifold, $M_n$, the operator $P$ is uniquely defined up to a 
factor $N$ which satisfies $N^2=I$ and commutes with all the operators of the 
spin theory on $M_n$. In examples we know this is $N=\pm I$ (see the Appendix
C). Thus $P$ is in some sense independent on the discrete symmetry group
where is involved, playing a similar role as the matrix $\gamma^5$ in the
usual theory of the Dirac spinors. 

When the metric is pseudo-Euclidean then the operators of the spinor 
representation of the discrete group ${\Bbb Z}_4(f^*)$ produced by 
$f^*\not =f$ have to be written directly using the Dirac adjoint. From 
Eq. (\ref{DUDU}) we obtain 
\begin{equation}
U_{f^*}=(\overline{U_f})^{-1}=\overline{U_f}\,\,\overline{P}
\end{equation}
where, as observed, $\overline{P}=\pm P$. One can show that the sign is $(+)$   
when $\gamma$ is built from an even number of gamma matrices and $(-)$ when 
this number is odd. Starting with these elements  the remaining operators of 
the cyclic group will be obtained using multiplication. 

A most interesting case is that of the discrete symmetries of the Dirac-type  
operators $D^i$ ($i=1,2,3$) given by the triplet $F$ which satisfy Eqs. 
(\ref{algf}). 
\begin{theor}
The Dirac operators $\pm D$ and the Dirac-type ones 
$\pm D^1$, $\pm D^2$, and $\pm D^3$ are related among themselves through the 
transformations of the spinor representation of the quaternion group,  
${\Bbb Q}(F)\subset {\cal G}_F$.
\end{theor}  
\begin{demo}
Let us denote by $U_i=T(-\frac{\pi}{2}f^i)$ the operators that, according to 
Theorem \ref{D1}, have the properties 
\begin{equation}
(U_1)^2=(U_2)^2=(U_3)^2=P\,,\quad P^2=I\,,
\end{equation} 
and $PU_i=U_i P$. Furthermore, from Eqs. (\ref{TgT1}), (\ref{Lamaf}) and 
(\ref{algf}) we deduce 
\begin{eqnarray}
&U_1 U_2=U_3\,, \quad U_2 U_3=U_1\,, \quad &U_3 U_1=U_2\,,\\  
&U_2 U_1=U_3 P\,, \quad U_3 U_2=U_1 P\,, \quad &U_1 U_3=U_2 P\,,  
\end{eqnarray}
and, after a few manipulation, we see that  $I$, $P$, $U_i$ and 
$PU_i$ ($i=1,2,3$) form a representation of the dicyclic group 
$\left<2,2,2 \right>$ which is isomorphic with the quaternion subgroup of 
${\cal G}_F$ we denote by ${\Bbb Q}(F)$.
Using Eqs. (\ref{DUDU}) and (\ref{pari}) we find that 
its orbit in the space of the Dirac operators is the desired one. 
\end{demo}\\ 
As expected, the cyclic groups  ${\Bbb Z}_4(f^i)$ are subgroups of 
${\Bbb Q}(F)$. For this reason the spinor representation of the group 
${\Bbb Q}(F^*)$ has to be derived from that of ${\Bbb Q}(F)$ using the same 
method as in the case of cyclic groups. 

Hence we conclude that for each isolated unit root $f$ one can define a finite 
group ${\Bbb Z}_4(f)$ which is a subgroup of ${\cal G}_f$ while the triplets 
$F$ produce  more complicated finite discrete groups, ${\Bbb Q}(F)\subset 
{\cal G}_F$. Since the groups ${\cal G}_f$ and ${\cal G}_F$ can not be embedded 
in a larger group, the product of two operators of the spinor representation 
of two different discrete groups is, in general, an arbitrary operator which
do not correspond to a transformation of another discrete group ${\Bbb Z}_4$ 
or ${\Bbb Q}$. In addition, this new operator could transform the standard 
Dirac operator in a new operator having  different properties to those of the 
Dirac-type ones. Therefore when we restrict ourselves to orbits containing 
only the Dirac  and Dirac-type operators we have to consider only the discrete 
groups discussed above.   
 
\section{Concluding remarks}

Here we have studied the symmetries and  supersymmetries of the 
Dirac-type operators built with the help of the unit roots which in
general do not accomplish the complex structures of the K\" ahlerian 
geometries. 
This extension, that seems to be pointless in geometry and 
classical physics, is very useful from the point of view of the Dirac theory 
since it permits to construct Dirac-type operators even in geometries which do 
not admit complex structures, as in the case of the Minkowski spacetime.
Therefore, now it is certain that isolated or triplets of Dirac-type
operators arise in {\em any flat manifold}
of dimension $n=2k$ indifferent on the metric signature.    
Of course,  the triplets of Dirac-type operators produced by families of unit 
roots could appear only in  manifolds of dimensions $n=4k$.
 
On the other hand, the use of the roots in manifold of arbitrary dimensions 
brings nothing new from the point of view of the symmetry and 
supersymmetry of the Dirac-type operators since these can not be larger than 
those we have found, for example, in the Euclidean Taub-NUT space that is a 
hyper-K\" ahler manifold \cite{CV,CV1}. Thus in any concrete theory we expect 
to find at most continuous $SU(2)$ symmetries coupled to quaternionic discrete 
ones  associated to the $N=4$ superalgebras of Dirac and Dirac-type operators. 
What is remarkable is that there are geometries with many such superalgebras 
that can not be embedded in larger one showing off a higher symmetry. This 
is a new phenomenon which may have consequences not only in the Dirac theory 
but in geometry too. We hope that our method could be a good tool for 
investigating delicate problems concerning the topology of these rich sets of 
Dirac-type operators. 

Technically speaking, the important results we obtained here are the 
rule (\ref{DDS}) of generating Dirac-type operators and the definitive form of 
the transformations (\ref{TDT10}) and (\ref{TDT20}) of the representation 
$\sigma({\cal G}_F)$. In both cases the main role is played by the spin-like 
operators $\Sigma^i$ which simultaneously give rise to the Dirac-type 
operators  and define the generators of these transformations. However, this
fact is not surprising since one knows that they have  pseudo-classical
analogues with an interesting behavior.

In the pseudo-classical spinning particle models in curved spaces from 
covariantly constant K-Y tensors $f_{\mu\nu}$ can be constructed 
conserved quantities of the type $f_{\mu\nu} \theta^\mu \theta^\nu$ 
depending on the Grassmann variables $\{\theta^\mu\}$ \cite{VV}. 
The Grassmann variables $\{\theta^\mu\}$ transform as a tangent space 
vector and describe the spin of the particle. The antisymmetric tensor 
$S^{\mu\nu} = - i \theta^\mu \theta^\nu$ generates the internal part of the 
local tangent space rotations. For example, in the spinning Euclidean 
Taub-NUT space such operators correspond to components of the spin which 
are separately conserved \cite{JWH}.

The construction of the new supersymmetries in the context of
pseudo-classical mechanics can be carried over straightforwardly to the
case of quantum mechanics by the usual replacement of phase space
coordinates by operators and Poisson-Dirac brackets by anticommutators
\cite{BeMa}. In terms of these operators the supercharges are replaced by
Dirac-type operators \cite{Ri}. In both cases, the correspondence principle
leads to equivalent algebraic structures making obvious the relations
between these approaches \cite{JWH}.

It is worth to mention the role of non-covariantly constant K-Y tensors 
in generating hidden symmetries. The presence of non-covariantly 
constant K-Y tensors implies the existence of non-standard 
supersymmetries in point particle theories on curved background. In 
general, in such a case, the St\" ackel-Killing tensor $K_{\mu\nu}$ in 
Eq. (\ref{i1}) is not covariantly constant and the corresponding 
non-standard supercharges $Q_a$ of hidden supersymmetries do not close 
on the Hamiltonian \cite{VV} (or on $D^2$, as in standard Dirac theories),
but $\{ Q_a , Q_b \} \propto J \delta_{ab}$, where $J$ is an invariant
different from the Hamiltonian (or $D^2$).

The Dirac type operators associated with the non-covariantly constant 
K-Y tensors were investigated in the case of the hidden symmetries of 
the Euclidean Taub-NUT space \cite{CV}. 
A more systematic study of the symmetries generated by the 
non-co\-va\-ri\-ant\-ly constant K-Y tensors and the properties of the 
corresponding Dirac operators will be given elsewhere.

\subsection*{Acknowledgments} The authors would like to express their
gratitude to the anonymous referees for comments and suggestions that 
have improved this article.

\setcounter{equation}{0} \renewcommand{\theequation}
{A.\arabic{equation}}

\section*{Appendix A\\K\" ahlerian geometries}

Let us consider the manifold $M_n$ ($n=2k$) and its tangent fiber bundle, 
${\cal T}(M_{n})$,  assuming that $M_n$ is equipped with a 
{\em  complex structure} that is a particular bundle automorphism 
$h: {\cal T}(M_{n}) \to {\cal T}(M_{n})$ which satisfies 
$\left<h\right>^2=-{\bf I}$ and is covariantly constant. 
Notice that the matrix of $h$ in local frames is an orthogonal
point-dependent transformation of the gauge group $G(\eta)$.
With its help one gives the following definition \cite{LM,GM}: 
\begin{defin} 
A Riemannian metric $g$ on $M_n$ is said K\" ahlerian if $h$ is 
pointwise orthogonal, i.e., $g(hX,hY)=g(X,Y)$ for all $X,Y\in 
{\cal T}_x(M_{n})$ at all points $x$. 
\end{defin}  
In local coordinates, $h$ is a skew-symmetric second rank tensor with 
real-valued components, $h_{\mu\nu}=-h_{\nu\mu}$, 
which obey $g_{\mu\nu}h^{\mu\,\cdot}_{\cdot\,\alpha}
h^{\nu\,\cdot}_{\cdot\,\beta}=g_{\alpha\beta}$. This gives rise to the 
symplectic form $\tilde\omega=\frac{1}{2}h_{\nu\mu}dx^{\nu}\land dx^{\mu}$ 
(i.e., closed and non-degenerate). 
Alternative definitions can be formulated starting with both, $g$ and 
$\tilde\omega$, which have to satisfy the K\" ahler relation 
$\tilde\omega(X,Y)=g(X,hY)$ \cite{GM}.    

A {\em hypercomplex structure} on $M_n$ is an ordered triplet $H = (h^1, h^2, 
h^3)$ of complex structures on $M_n$ satisfying Eq. (\ref{algf}). 
In Lie algebraic terms, the matrices ${1 \over 2} \left<h^j\right>$ realize 
the $su(2)$ algebra.
\begin{defin}
A hyper-K\" ahler manifold is a manifold whose Riemannian metric is 
K\"ahlerian with respect to each different complex structures 
$h^1, h^2$ and $h^3$. 
\end{defin}

Our unit roots, $f$, are defined in a similar way as the complex structures 
with the difference that the unit roots are automorphisms of the 
{\em complexified} tangent bundle, 
$f: {\cal T}(M_{n})\otimes{\Bbb C} \to {\cal T}(M_{n})\otimes {\Bbb C}$. 
Therefore, $f$ have complex-valued components and the transformation 
matrix $\left<f\right>$ is of the complexified group $G_{c}(\eta)$. Thus 
it is clear that the real-valued unit roots are complex structures as defined 
above. The families of unit roots may differ from the hypercomplex structures 
but have the same algebraic properties given by Ec. (\ref{algf}). 

The passing from the complex structures to unit roots is productive 
from the point of view of the Dirac theory since in this way 
one can introduce families of unit roots generating superalgebras of 
Dirac-type operators even in manifolds which do not admit complex structures.   
The Minkowski spacetime is a typical example. This will be briefly presented 
after discussing the Euclidean Taub-NUT space which is hyper-K\" ahler
\cite{AH, GR}.  

\setcounter{equation}{0} \renewcommand{\theequation}
{B.\arabic{equation}}

\section*{Appendix B\\Example: Euclidean Taub-NUT space}

In the spacetime of the Kaluza-Klein theory, $M_{(1,4)}$, corresponding to the 
usual Dirac spinor space, the Euclidean Taub-NUT space, $M_{(,4)}$, covers the 
whole space part which is of dimension four. We take the metric 
$\eta={\rm diag}(-1,-1,-1,-1)$ and a 
chart with Cartesian coordinates $x^{\mu}$  ($\mu, \nu,...=1,2,3,5$) where the 
line element reads 
\begin{equation}\label{(met)} 
ds^{2}=g_{\mu\nu}dx^{\mu}dx^{\nu}=-\frac{1}{V}dl^{2}-V(dx^{5}+
A_{i}dx^{i})^{2}\,.
\end{equation}   
The notation $dl^{2}=(d\vec{x})^{2}$ stands for the  Euclidean 
three-di\-men\-sio\-nal line element and $\vec{A}$ is the gauge field of a 
monopole. Another chart suitable for applications is 
of spherical coordinates, $(r,\,\theta,\,\phi,\,\chi)$, among them 
the first three are the  spherical coordinates associated with 
the  Cartesian space ones, $x^{i}$  ($i,j,...=1,2,3$), and 
$\chi+\phi=- x^{5}/\mu$. The real number $\mu$ is the  parameter  
defining the function $1/V(r)=1+\mu/r$ while the unique non-vanishing 
component of the vector potential in spherical coordinates is 
$A_{\phi}=\mu(1-\cos\theta)$. 

Furthermore, we consider the local frames given by tetrad fields 
$e(x)$ and $\hat e(x)$ as defined in \cite{P} while the four Dirac matrices 
$\gamma^{\hat\alpha}$, that  satisfy 
$\{ \gamma^{\hat\alpha},\, \gamma^{\hat\beta} \} 
=-2\delta^{\hat\alpha \hat\beta}$, have to be written in the following  
representation 
\begin{equation}\label{(gammai)} 
\gamma^i = 
\left(
\begin{array}{cc}
0&\sigma_i\\
-\sigma_i&0
\end{array}\right)\,,\quad  
\gamma^5 =i
\left(
\begin{array}{cc}
0&{\bf 1}_2\\
{\bf 1}_2&0
\end{array}\right)\,.
\end{equation}
In addition we consider the matrix
$\gamma^0 = \gamma^1\gamma^2\gamma^3\gamma^5 ={\rm diag}({\bf 1}_2,
-{\bf 1}_2)$
which is involved in the Kaluza-Klein theory of The Dirac equation explicitly 
depending on time \cite{CV0}.

The {\it standard} Dirac operator of the theory without explicit mass term 
is defined in local frames as 
${D}=i\gamma^{\hat\alpha}\nabla_{\hat\alpha}$  \cite{CV0,CV1}
where the spin covariant derivatives with local indices, 
$\nabla_{\hat\alpha}$, depend on the momentum operators,  
$P_{i}=-i(\partial_{i}-A_{i}\partial_{5})$ and $P_{5}=-i\partial_{5}$,
and spin connection \cite{CV0},  such that the  Dirac operator 
\begin{equation}\label{HH}
D =\left(
\begin{array}{cc}
0&D^{(+)}\\
D^{(-)}&0
\end{array}\right)
\end{equation}
can be expressed in terms of Pauli operators,  
\begin{equation}
D^{(-)}
=\sqrt{V}\left(- \vec{\sigma}\cdot\vec{P}+\frac{iP_{5}}{V}\right)\,,\quad  
D^{(+)} 
=V\left(\vec{\sigma}\cdot\vec{P}+\frac{iP_{5}}{V}\right)\frac{1}{\sqrt{V}}\,, 
\end{equation}
involving the Pauli matrices, $\sigma_i$. Notice that the Klein-Gordon operator 
of the Euclidean Taub-NUT space  $\Delta= \nabla_{\mu}g^{\mu\nu}\nabla_{\nu}=
-D^{(+)}D^{(-)}$ commutes with $\sigma_i$ but  $D^{(-)}D^{(+)}$ does not have 
this property because of its complicated form involving spin terms 
\cite{CV0,CV}. 

For this reason there is only one triplet of unit roots, $F=\{f^{(1)}, f^{(2)}, f^{(3)}\}$,   
which satisfy Ec. (\ref{algf}). Their non-vanishing components in the local 
frame are real,
\begin{eqnarray}
\hat f^{(1)}_{23}=1\,,\quad&\hat f^{(2)}_{31}=1\,,&\quad
\hat f^{(3)}_{12}=1\,,\\
\hat f^{(1)}_{51}=1\,,\quad&\hat f^{(2)}_{52}=1\,,&\quad
\hat f^{(3)}_{53}=1\,,
\end{eqnarray}
such that they form a hypercomplex structure and the Euclidean Taub-NUT space 
is hyper-K\"ahler. They give rise to the spin-like operators  
\begin{equation}\label{sl}
\Sigma^{(i)}=\frac{i}{4}\hat f^{(i)}_{\hat\alpha\hat\beta}
\gamma^{\hat\alpha}\gamma^{\hat\beta}=\left(
\begin{array}{cc}
\sigma_i&0\\
0&0
\end{array}\right)\,,
\end{equation}
and, according to Eq. (\ref{DDS}), produce the  Dirac-type operators 
\cite{CV0}
\begin{equation}\label{Dto} 
D^{(i)}=i[D,\,\Sigma^{(i)}]=\left(
\begin{array}{cc}
0&-i\sigma_i D^{(+)}\\
iD^{(-)}\sigma_i&0
\end{array}\right)
\end{equation}
which anticommute with $D$ and $\gamma^0$. 

The $SU(2)$ transformations (\ref{TxiS})  
generated  by the above defined spin-like operators are 
\begin{equation}\label{Uxi}
T(\vec{\xi})=e^{i\vec{\xi}\cdot \vec{\Sigma}}=\left(
\begin{array}{cc}
\hat U(\vec{\xi})&0\\
0&{\bf 1}_2
\end{array}\right)\,,
\end{equation}
where $\vec{\xi}=\alpha \vec{\nu}$ ( $\vec{\nu}^2=1$) and $U$ is the $SU(2)$ 
transformation 
\begin{equation}\label{SU2} 
\hat U(\vec{\xi})=e^{i\vec{\xi}\cdot\vec{\sigma}}=
{\bf 1}_2 \cos\alpha +i\vec{\nu}\cdot\vec{\sigma}\sin\alpha\,.
\end{equation}
The last step is to write down the final form of the concrete 
transformations according to Eqs. (\ref{TDT10}) and (\ref{TDT20}).

The discrete transformations of the group ${\Bbb Q}(F)$ are 
$I$,  $P={\rm diag}(-{\bf 1}_2,{\bf 1}_2)$ and the  sets of matrices  
$U_{(k)}={\rm diag} (-i\sigma_{k},{\bf 1}_2)$ and $PU_{(k)}$.

In the Euclidean Taub-NUT space, in addition to the above discussed 
complex structure, there exists a fourth 
Killing-Yano tensor,  
\begin{equation}\label{fy}
f^Y_{\alpha\beta} 
= - \frac{x^i}{r}f^{(i)}_{\alpha\beta} +\frac{2 x^i}{\mu V}\varepsilon_{ijk} 
\,{\hat e}^j_{\alpha} 
\,{\hat e}^k_{\beta}\,,
\end{equation}
which is not covariantly constant. The presence of $f^Y$ is due to the 
existence of the hidden symmetries of the Euclidean Taub-NUT geometry which are 
encapsulated in three non-trivial St\"ackel-Killing tensors. These 
are interpreted as the components of the so-called Runge-Lenz vector of 
the Euclidean Taub-NUT problem and are expressed as symmetrized products of
the Killing-Yano tensors $f^Y$ and $f^{(i)},\, (i=1,2,3)$ \cite{CV}.

\setcounter{equation}{0} \renewcommand{\theequation}
{C.\arabic{equation}}

\section*{Appendix C\\Example: Minkowski spacetime}

In the Minkowski spacetime, $M_{(1,3)}\subset M_{(1,4)}$, with 
metric $\eta=(1,-1,-1,-1)$ we take the gauge 
$e^{\mu}_{\nu}=\hat e^{\mu}_{\nu}=\delta^{\mu}_{\nu}$, ($\mu,\nu,...=0,1,2,3$). 
We use the chiral representation of the Dirac matrices where the standard 
Dirac operator reads
\begin{equation}
D=i\gamma^{\mu}\partial_{\mu}=\left( 
\begin{array}{cc}
0&i(\partial_t+\vec{\sigma}\cdot\vec{\partial})\\
i(\partial_t-\vec{\sigma}\cdot\vec{\partial})&0
\end{array}\right)=\left(
\begin{array}{cc}
0&D^{(+)}\\
D^{(-)}&0
\end{array}\right)
\end{equation}
and the generators of the spinor representation of the group 
${\cal G}(\eta) = SL(2,{\Bbb C})$ take the form
$S^{ij}=\frac{1}{2}\varepsilon_{ijk}{\rm diag}(\sigma_k, \sigma_k)$
and $S^{i0}=\frac{i}{2}{\rm diag}(\sigma_i,-\sigma_i)$.

There are two families of three roots \cite{K2}. The unit roots of the first 
triplet, $F$, have the non-vanishing complex components \cite{K2}
\begin{eqnarray}
f^{(1)}_{23}=1\,,\quad&f^{(2)}_{31}=1\,,&\quad f^{(3)}_{12}=1\,,\\
f^{(1)}_{01}=i\,,\quad&f^{(2)}_{02}=i\,,&\quad f^{(3)}_{03}=i\,,
\end{eqnarray}
and, consequently, $F$ is not a hypercomplex structure and the Minkowski 
spacetime is not a hyper-K\" ahler manifold.
The corresponding spin-like operators
\begin{equation}
\Sigma^{(i)}=\frac{1}{2}f^{(i)}_{\mu\nu}S^{\mu\nu}=\left(
\begin{array}{cc}
\sigma_i&0\\
0&0
\end{array}\right)
\end{equation}
give Dirac-type operators $D^{(i)}=i[D,\,\Sigma^{(i)}]$ of the same form as 
(\ref{Dto})
and the generators ${\cal S}^{(i)}=\frac{1}{2}\Sigma^{(i)}$ of the spinor 
representation of the group ${\cal G}_F=SU(2)$. 
The second triplet is $F^*$ for which all the spinor quantities are 
just the Dirac conjugated of those of $F$ (using $\gamma=\gamma^0$). 
The corresponding spin-like operators are   
\begin{equation}
\overline{\Sigma^{(i)}}=\frac{1}{2}\left(f^{(i)}_{\mu\nu}\right)^{*}S^{\mu\nu}
=\left(
\begin{array}{cc}
0&0\\
0&\sigma_i
\end{array}\right)
\end{equation}
which means that the $SU(2)$ groups $\sigma({\cal G}_F)$ and 
$\sigma({\cal G}_{F^*})$ act {\em separately} on the left and right-handed 
parts of the Dirac spinor. Moreover, it is interesting to observe that 
$\Sigma^{(i)}+\overline{\Sigma^{(i)}}=2 S^i$  where 
$S^i=\frac{1}{2}\varepsilon_{ijk}S^{jk}$ are the usual spin operators.
This perfect balance between the chiral sectors is due to the fact that 
the operator $D^{(+)}D^{(-)}=D^{(-)}D^{(+)}$ commutes with $\sigma_i$.

The discrete symmetry is given by two representations of the quaternion group 
acting on each of both chiral sectors. On the left-handed sector acts the 
group  ${\Bbb Q}(F)$ represented by the operators $I$, 
$P={\rm diag}(-{\bf 1}_2,{\bf 1}_2)$, 
$U_{(i)}={\rm diag}(-i\sigma_i, {\bf 1}_2)$ and  $P U_{(i)}$. The operators 
of the group of the right-handed sector, ${\Bbb Q}(F^*)$, can be 
obtained using the Dirac adjoint and taking into account that here 
$\overline{P}=-P$. The resulted operators, 
${\rm diag}({\bf 1}_2, \pm i\sigma_i)$, 
act only on the right-handed sector. 

Hence it is clear that each chiral sector has its own sets of unit roots 
defining Dirac-type operators. These are  the spheres ${\bf S}_F$ of the 
left-handed sector and ${\bf S}_{F^*}$ of the right-handed one. Since there are 
no other independent unit roots, the whole set of unit roots  of the Minkowski 
spacetime is ${\cal R}_1(M)={\bf S}_{F}\bigcup {\bf S}_{F^*}$. The continuous 
and discrete symmetry groups of the Dirac-type operators are defined separately 
on each of these two spheres. The continuous one is governed by two separated 
Lie algebras, $L_F\sim so(3)$ (for the left-handed sector) and 
$L_{F^*}\sim so(3)$ (for the right-handed one), rather than by a total Lie 
algebra $so(3)\oplus so(3)\sim so(4)$ \cite{K2}.

\end{document}